\documentstyle[11pt,newpasp,twoside]{article}
\def\edcomment#1{\iffalse\marginpar{\raggedright\sl#1\/}\else\relax\fi}
\reversemarginpar

\begin{document}
\title{Visualising Interacting Binaries in 3D} 

\author{R. I. Hynes} 

\affil{Department of Physics and Astronomy, University of
Southampton, Southampton, SO17 1BJ, UK; rih@astro.soton.ac.uk}

\begin{abstract}
I have developed a code which allows images to be produced of a
variety of interacting binaries for any system parameters.  The
resulting images are not only helpful in visualising the geometry of a
given system but are also helpful in talks and educational work.
\end{abstract}

Many artists impressions of interacting binaries circulate.  These
have limitations; each only represents one system, at one phase and
sometimes non-physical mass ratios, temperatures, etc., are used.
Some codes do exist which produce wireframe images and these are often
seen in papers.  To complement these I have developed a code, {\sc
BinSim}, to produce color images for any set of parameters.  It is
useful for visualising the geometry of the system you are working on
and preparing images and movies for conference talks, lectures,
educational work, press releases, etc.  The emphasis is on attractive
pictures rather than strict scientific accuracy, and there are many
simplifications, but the aim has been not to stray too far from our
current scientific view of these objects.

Various kinds of binaries can be represented, including disc accreting
CVs (dwarf novae, novalikes), low mass X-ray binaries, some high mass
X-ray binaries and detached white dwarf M dwarf binaries.  To do this,
a configurable selection of building blocks are used: a companion star
(lobe filling or not), an optically thick or thin accretion disc, a
stream and hot-spot, a disc corona or wind and a jet.  Images can be
saved as JPEGs and movies can also be made, either in real time or to
MPEG files.  The latter include orbital motion, Keplerian disc
rotation and motion along the stream.

The basic approach is to break the binary up into distinct components:
the companion, disc, stream, hotspot, etc.  Then each component is
divided into a rectangular grid of vertices.  This grid defines
triangles covering the surface.  Colours within a triangle are
interpolated between the vertices.  All components can be represented
by such a grid, for example wrapping it into a curved cylinder for the
stream, wrapping into a sphere for the hot-spot, and a distorted
sphere for the companion.  A flattened sphere with the poles joined
represents the disc This approach allows common code between
components, hence adding new components is relatively easy.  All
patterns are generated in real time for maximum flexibility, so there
is no need for pregenerated texture maps.

For optically thick objects, e.g.\ the companion star, it is just
necessary to define the temperature and normal vector at each vertex.
`Red', `green' and `blue' fluxes can be calculated using a black body
spectrum and a linear limb-darkening law.  If actual red/green/blue
wavelengths used, then the colours look washed out into shades of
brown.  {\sc BinSim} emphasises the colour differences by actually
using IR/optical/UV wavelengths.  Fluxes are scaled logarithmically
(as the eye does) to get RGB colour values.  A reference temperature,
7000K, is taken to be white.  Hotter objects then appear blue, cooler
ones are orange or red.  Contrast and brightness adjustments can also
be made at this stage.

Components which are wholly or partially optically thin present extra
problems.  At present {\sc BinSim} models them as partially
transparent surfaces; this is the most natural approach when
constructing objects from triangles.  Unfortunately such objects are
hollow and internal structure must be represented on the surface.
Some objects are not readily modelled in this way, e.g.\ `fluffy'
discs, and in animations this representation is visibly incorrect.  A
better approach would be to use fully volumetric models with their
internal structure defined in 3D.  One way to do this is, for each
phase, to ray trace back from the surface to calculate projected
optical depth at each point.  Another is to model the object as 3D
grid of volumetric elements (voxels).  Either approach will be much
more demanding in CPU and/or memory, so may only be practical for
stills.

There are many things planned or possible with future versions of {\sc
BinSim}.  Support for extra binary components would increase the range
of systems that could be modelled, for example a non-compact primary
would allow Algols and non-interacting but deformed OB binaries;
magnetic structures could be added for magnetic CVs and circumstellar
discs would allow Be X-ray binaries to be represented.  A more general
deformable disc model would allow stream-impact bulges, warped or
elliptical discs and spiral shocks.  Fully 3D volumetric models for
optically thin objects could represent `fluffy' discs, non-uniform
winds and a more realistic hot-spot.  Further animation support could
include a flickering irradiation source and irregular jet ejections.
Support could be added for non-binary models e.g.\ AGN and protostars.

The C++ code can be downloaded from the {\sc BinSim} webpage
\footnote{\texttt{http://\-www.astro.soton.ac.uk/\-$\sim$rih/\-binsim.html}};
example images are also available here.  The code should be usable on
any Unix platform but has been mainly tested on Linux.  At least three
other codes exist, by Dan Rolfe, Andy Beardmore and Jens Kube.  All
three do things that the others do not, so they are partly
complementary.

\bigskip
\acknowledgements{I would like to acknowledge financial support from
the Leverhulme Trust, and to thank Dan Rolfe for many discussions on
how to represent interacting binaries and the users of {\sc BinSim}
who have provided valuable testing and feedback.  {\sc BinSim} would
not have been possible without the efforts of Brian Paul and others
responsible for the Mesa 3-D graphics
library\footnote{http://www.mesa3d.org/}.}

\end{document}